\documentclass[aip,amsmath,amssymb,reprint]{revtex4-1}
\draft
\usepackage{graphicx}  % needed for figures
\usepackage{dcolumn}% Align table columns on decimal point
\usepackage{amsmath}
\usepackage{makecell}
\usepackage{bm}        % for math
\usepackage{amssymb}   % for math

\usepackage[utf8]{inputenc}
\usepackage[T1]{fontenc}
\usepackage{mathptmx}

\usepackage{float}
\usepackage{color}

\begin{document}

\title{Comparing machine learning techniques for predicting glassy dynamics
}

%Machine learning

\author{Rinske M. Alkemade}
\affiliation{Soft Condensed Matter, Debye Institute of Nanomaterials Science, Utrecht University, Utrecht, Netherlands }
\author{Emanuele Boattini}
\affiliation{Soft Condensed Matter, Debye Institute of Nanomaterials Science, Utrecht University, Utrecht, Netherlands }
\author{Laura Filion}
\affiliation{Soft Condensed Matter, Debye Institute of Nanomaterials Science, Utrecht University, Utrecht, Netherlands }
\author{Frank Smallenburg}
\affiliation{
Universit\'e Paris-Saclay, CNRS, Laboratoire de Physique des Solides, 91405 Orsay, France
}
\begin{abstract}
In the quest to understand  how structure and dynamics are connected in glasses, a number of machine learning based methods have been developed that predict dynamics in supercooled liquids. These methods include both increasingly complex machine learning techniques, and increasingly sophisticated descriptors used to describe the environment around particles. In many cases, both the chosen machine learning technique and choice of structural descriptors are varied simultaneously, making it hard to quantitatively compare the performance of different machine learning approaches. Here, we use three different machine learning algorithms -- linear regression, neural networks, and GNNs -- to predict the dynamic propensity of a glassy binary hard-sphere mixture using as structural input a recursive set of order parameters recently introduced by Boattini {\it et al.} [Phys. Rev. Lett. {\bf 127}, 088007 (2021)].  As we show, when these advanced descriptors are used, all three methods predict the dynamics with nearly equal accuracy. However, the linear regression is orders of magnitude faster to train making it by far the method of choice. 
\end{abstract}

\maketitle

The relationship between local structure and dynamics in glassy systems has been a heavily debated question in condensed matter for several decades \cite{theoryglasses, patty, tanaka2019revealing}. Over the last few years, one avenue for exploring this relationship has been an effort to predict dynamical behavior based on local structural features using various machine learning (ML) methods. Pioneered by Cubuk {\it et al.}\cite{structurevsdynamics1} using support vector machines (SVMs) to predict rearrangement probabilities in glassy mixtures, this area of research has now embraced a wide variety of ML techniques, including e.g. linear regression, convolutional neural networks (CNNs), graph neural networks (GNNs), autoencoders, and community inference, see e.g. Refs. \onlinecite{BAPST, glassydnamics2, structurevsdynamics5, PaperEmanuele, andrea, richard2020predicting}. This raises the question of what ML technique one should choose when predicting the dynamics of a glassy system.

This question is far from straightforward, since in addition to choosing a machine learning technique, one also has to make a choice with respect to the encoding of the local structure in terms of data that can be interpreted by an ML algorithm. For most ML approaches, the structure around a particle is encoded into a set of structural order parameters that capture e.g. the local density and symmetry of the distribution of neighbors around that particle. However, some sophisticated ML approaches can work from much more restricted data. For example, in Ref. \onlinecite{BAPST}, it was shown that graph neural networks are capable of predicting the dynamic propensity of a glassy Lennard-Jones mixture based solely on encoding the structure into a graph of nearest neighbors and the pair distances between them. Such data would not be sufficient for e.g. a simple linear regression approach, but using a GNN  it was enough to drastically outperform both SVMs and CNNs trained with more sophisticated input. More recently, however, it was shown that even a simple linear regression approach can rival the predictive power of GNNs if supplied with sufficiently intelligent input data \cite{PaperEmanuele}. In particular, Boattini {\it et al.} proposed a method to iteratively construct generations of structural order parameters that successively take into account locally averaged order in expanding shells. These descriptors, in combination with linear regression preformed essentially as well as a GNN which was fed just the particle coordinates. 

This raises the intriguing question of whether more sophisticated ML techniques supplied with more intelligently chosen structural parameters can result in even better predictions. Here, we use three different ML algorithms -- linear regression, neural networks, and GNNs -- to predict the dynamic propensity of a glassy binary hard-sphere mixture,  based on the hierarchical set of order parameters from Ref. \onlinecite{PaperEmanuele}, and compare and contrast the results. As we show, out of the three methods, linear regression provides the best compromise between accuracy and efficiency when combined with these advanced structural descriptors.

\section{Model and descriptors}

\subsection{Model}

The glassy system that we use here to compare the three different ML methods is a binary hard-sphere mixture at packing fraction $\eta = 0.58$. It consists of hard spheres of two sizes, with a size ratio of $\sigma_S/\sigma_L= 0.85$, where $\sigma_{L(S)}$ is the diameter of a large (small) particle. The composition $x_L = N_L / (N_L + N_S) = 0.3$, where $N_{L(S)}$ is the number of large (small) spheres.  Note that this is the same glassy mixture as was studied in Refs. \onlinecite{Tetrahedrality, glassydnamics2, PaperEmanuele}. 

We simulate the evolution of our system using event-driven molecular dynamics (EDMD) \cite{EDMD}. The simulations are performed in the microcanonical ensemble (constant number of large and small particles $N_L$ and $N_S$, volume $V$, and kinetic energy $U$). The time unit of our simulation is defined as $\tau = \sqrt{m\sigma_L^2/k_B T}$ where $k_B$ is Boltzmann's constant and $m$ is the particle mass. Note that we set the masses of all particles to be equal. All simulated systems contained 2000 particles in total (600 large, 1400 small).

To generate the initial configurations, we use a separate EDMD simulation in which the particles grow over time until the desired packing fraction $\eta = 0.58$ is reached. After this, the system is equilibrated for at least $10^5 \tau$. Note that from previous work \cite{glassydnamics2}, we know that the relaxation time of this system is on the order of $\tau_\alpha = 10^4 \tau$.

\subsection{Dynamic propensity}

To characterize the dynamical heterogeneity in our glassy system, we use the dynamic propensity \cite{dynamicpropensity, propensity1}. This quantity is closely related to the mean squared displacement, and captures how far individual particles on average move over time. To measure the dynamic propensity, the evolution of a glassy system is measured multiple times, each time starting from the same initial configuration, while assigning each particle a random velocity drawn from a Maxwell-Boltzmann distribution at the desired temperature. This ensemble is called the isoconfigurational ensemble\cite{dynamicpropensity}. To obtain the propensity $ \Delta r_i(t) $ of particle \textit{i}, we average the absolute distance it traveled over the time interval $t$ over all trajectories
\begin{equation}
    \Delta r_i(t) = \left\langle |\mathbf{r}_i(t) - \mathbf{r}_i(0)| \right\rangle_\mathrm{iso},
\end{equation}
where $\langle \cdot \rangle_\mathrm{iso}$ indicates the average taken over all trajectories in the isoconfigurational ensemble. 

To measure the propensity, we average over simulations starting from 100 different initial snapshots, and for each initial snapshot we average over 50 trajectories with different initial velocities. The dynamic propensity is measured at a logarithmically spaced set of time intervals $t$, between $t/\tau = 10^{-2}$ and $t/\tau = 10^5$.

\begin{figure}
    \centering
    \includegraphics[width=0.49\textwidth]{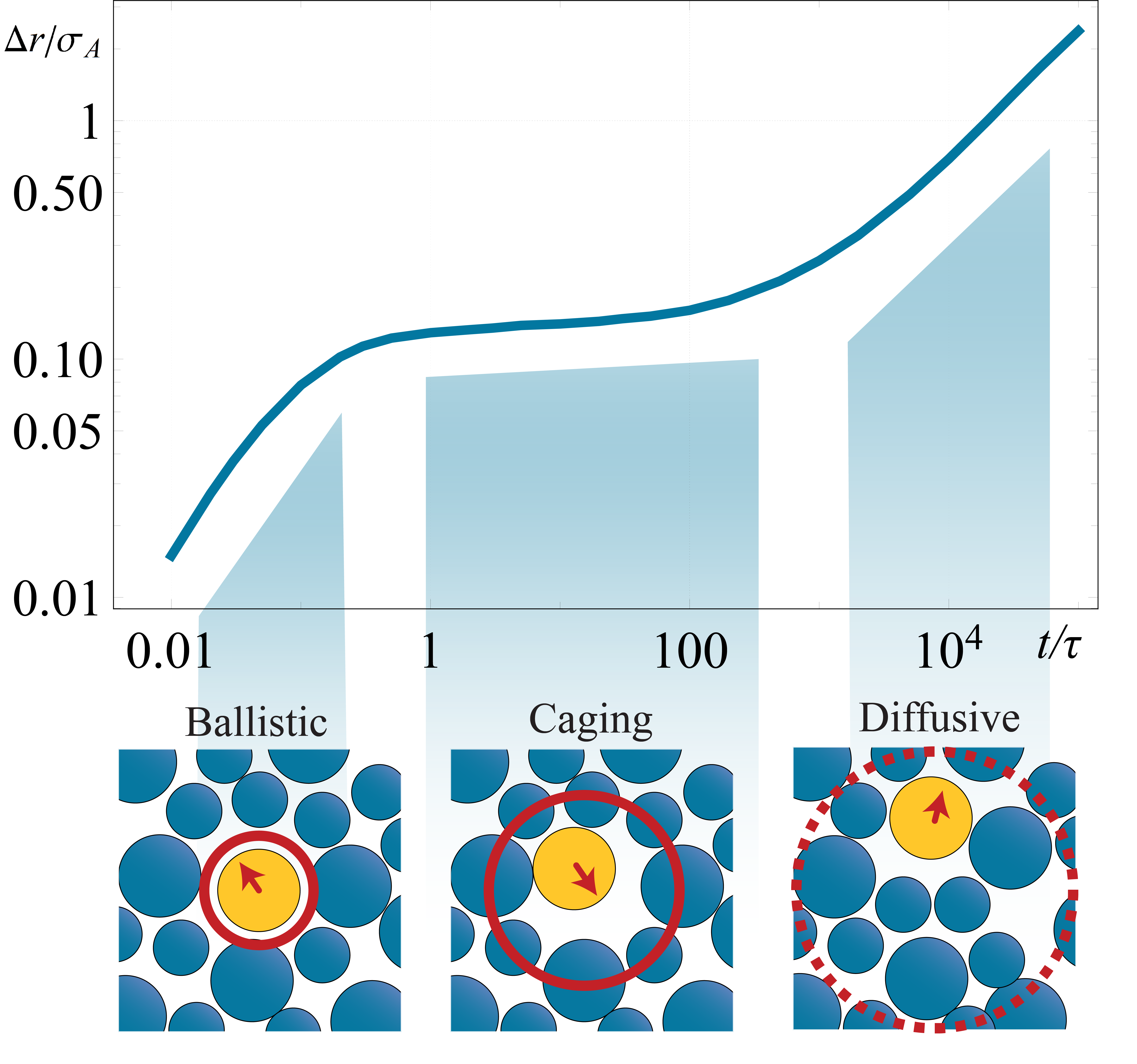}
    \caption{Globally averaged dynamical propensity $\langle\Delta r_i(t) \rangle$ as a function of time. The insets at the bottom illustrate three different dynamical regimes: the ballistic regime, in which particles are not yet interacting with their neighbors, the caging regime, in which particles are trapped by surrounding particles, and the diffusive regime, where particles have escaped their cages.}
    \label{fig:prop}
\end{figure}

To illustrate the behavior of the dynamic propensity, we plot in Fig. \ref{fig:prop} the globally averaged dynamic propensity as a function of time.

\subsection{Structural descriptors}

To describe the local environments of particles, we use the structural order parameters used in Ref.~\onlinecite{PaperEmanuele}, which consist of a combination of radial densities and angular functions measured in different shells around a particle. 

For the radial functions, we use essentially the same descriptors as used in Refs. \onlinecite{andrea, BAPST, PaperEmanuele}. These descriptors measure a weighted particle density inside a spherical shell with a thickness of approximately $2\delta$ at distance \textit{r} with respect to a reference particle $i$. The functions are defined as
\begin{equation}
    G^{(0)}_i(r, \delta, s) = \sum_{j\neq i, s_j = s}e^{-\frac{(r_{ij} -r)^2}{2\delta^2}}.
\end{equation}
 Here \textit{i} is the reference particle, $s$ is the particle type and $r_{ij}$ is the distance between particles \textit{i} and \textit{j}. The summation is carried out over all particles with particle type $s_j = s$, which means that the radial density that is measured is type-specific. 

The angular descriptors that we use are based on bond order parameters \cite{BOPS, BOPSold}. These bond order parameters expand the local environment in terms of spherical harmonics. To obtain the angular descriptors for a particle $i$, we first calculate the complex coefficients
\begin{equation}
    q_i^{(0)}(l, m, r, \delta) = \frac{1}{Z} \sum_{i\neq j} e^{-\frac{(r_{ij} -r)^2}{2\delta^2}}Y^m_l(\mathbf{r}_{ij}). 
\end{equation}
Here $Y^m_l(\mathbf{r}_{ij})$ is the spherical harmonic of order \textit{l}, with \textit{m} an integer that runs from $-l$ to $l$, and \textit{Z} is a normalization factor given by 
 \begin{equation}
 Z = \sum_{i\neq j} e^{-\frac{(r_{ij} -r)^2}{2\delta^2}}.
 \end{equation}
Note that although the summation runs over all particles, again the exponent makes sure that mainly particles within a spherical shell at distance $r$  and thickness $2\delta$ will contribute to $q_i^{(0)}(l, m, r, \delta)$.
Finally we sum over $m$ to obtain the rotationally invariant angular descriptors
\begin{equation}
    q_i^{(0)}(l, r, \delta) = \sqrt{\frac{4\pi}{2l+1}\sum_{m =-l}^{m =l}|q_i^{(0)}(l, m, r, \delta)|^2}.
\end{equation}
Due to the symmetries of the spherical harmonics,  $q^{(0)}(l, r, \delta)$ for a certain $l$ is expected to detect $l$-fold symmetry in the environment at the chosen distance $r$.

Boattini \textit{et al.} \cite{PaperEmanuele} showed  that the propensity prediction of a particle improves significantly, when the prediction is based not only on the structural parameters associated with the particle itself, but also on averaged structural information of neighbouring particles. Inspired by the architecture of graph neural networks, this was done by recursively constructing  higher-order averaged structural parameters, which are are defined as 
\begin{equation}
    x^{(n)}_i= \frac{\sum_{j: r_{ij}< r_c} x_j^{(n-1)} e^{-r_{ij}/r_c} }{\sum_{j: r_{ij}< r_c} e^{-r_{ij}/r_c}}, 
    \label{eq:higherorderbops}
\end{equation}
where $x_i$ can be any of the radial or angular order parameters of particle $i$. Additionally, $x^{(n)}_i$ represents the $n^\text{th}$ generation of parameter $x_i$, and the sum runs over all neighboring particles within a cutoff distance $r_c$, including $i$ itself.  The cutoff value $r_c$ is chosen to be $r_c/\sigma_L = 2.1$, which approximately corresponds to the second minimum of the radial distribution function \cite{PaperEmanuele}. However, as already shown in Ref.~\onlinecite{PaperEmanuele}, the exact value does not have a significant influence on our results. 

In total, we consider 354 0th-generation structural descriptors: 162 radial descriptors and 192 angular descriptors. 
For the radial descriptors we use 46 equally spaced spherical shells in the interval $r/\sigma_L = [0.86, 2.0]$, 20 equally spaced spherical shells in the interval $r/\sigma_L = (2.0, 3.0]$ and 15 equally spaced spherical shells in the interval $r/\sigma_L = (3.0, 4.5]$. For the angular descriptors we consider $l = 1$ to $12$ in 16 equally spaced spherical shells in the interval $[1,2.5]$. The full environment of each particle is then described with up to 3 generations of these 354 parameters each, leading to a total of 1062 parameters. \\

Before using the structural parameters as an input for the machine learning algorithms, they are standardized by evaluating
\begin{equation}
    \mathbf{x}^\text{st}_i = \frac{\mathbf{x}_i-\bar{\mathbf{x}}}{\mathbf{\sigma}_\mathbf{x}},
\end{equation}
where $\mathbf{x}_i$ is the vector containing all parameters associated with particle \textit{i}, $\mathbf{x}^\text{st}_i$ is the standardized parameter vector, and where $\bar{\mathbf{x}}$ and $\mathbf{\sigma}_\mathbf{x}$ are respectively the mean and standard deviation of the parameter vector considering all particles of the same species as $i$. The standardization ensures that all descriptors have zero mean and unit variance, which can be helpful when using regularization in machine learning techniques. This will be discussed in more detail below.

% The need to standardize our parameters arises due to the penalty that we add to large weights to prevent overfitting. Due to this penalty, providing the algorithm with not-standardized data would lend more weight to the parameters in the fit that have a large variance, since these parameters can have a large impact with a relatively low weight. For the same reasons that we standardize our input parameters, we also standardize the dynamic propensities for each time step separately.\\

% To train the machine learning algorithms, half of the obtained data is used as training data, while the other half is used as test data. For each time interval over which we measured the propensity, the models are trained with several different choices for the hyperparameters. Afterwards, the hyperparameter set that yields the highest Pearson correlation coefficient is chosen. Below we will discuss the different hyperparameters settings for each of the algorithms.

\section{Machine learning methods}

In this paper, we compare three different machine learning approaches for predicting the dynamic propensity based on the structural parameters introduced above. In particular, we compare linear regression (LR), neural networks (NN), and graph neural networks (GNN). Unless otherwise specified, we train separate models for large and small particles, and separate models for each time interval at which we are trying to predict the dynamic propensity.

Note that each of these approaches has a number of hyperparameters that tune the model fitted by the method to the supplied training data. For example, this can be the number of layers inside the neural network, parameters controlling regularization techniques that reduce overfitting, or the learning rate of the optimization algorithm for NNs and GNNs. 

\subsection{Linear regression}
Linear regression is the simplest of the three methods, and simply finds the best linear combination of all structural descriptors to predict the dynamic propensity. Here, we make use of L2-regularization (also known as Ridge regression) to reduce overfitting\cite{bookmachinelearning}. This approach penalizes large weights in the linear fit. Note that this is the reason we standardized our structural parameters: since the different parameters have the same mean and variance, the effect of the regularization on each parameter is the same.
For linear regression the only hyperparameter that can be tuned is $\alpha$, which sets the strength of the large-weight penalty in Ridge regression.

%For each time step we train the algorithm 10 times, each time with different values of $\alpha$ in a range of multiple orders of magnitude, i.e.  $\alpha \in [10^{-4}, 10^5]$. We furthermore provide the algorithm with different generations of parameters, i.e. (1) only the zeroth, (2) both the zeroth and the first, or (3) the zeroth, first and second generation. The regression is performed separately for small and large particles. 

\subsection{Neural network}
% In contrast to linear regression, an artifical neural network fits a non-linear model to the supplied training data. 

\begin{figure}
\centering
  \includegraphics[width=0.45\textwidth]{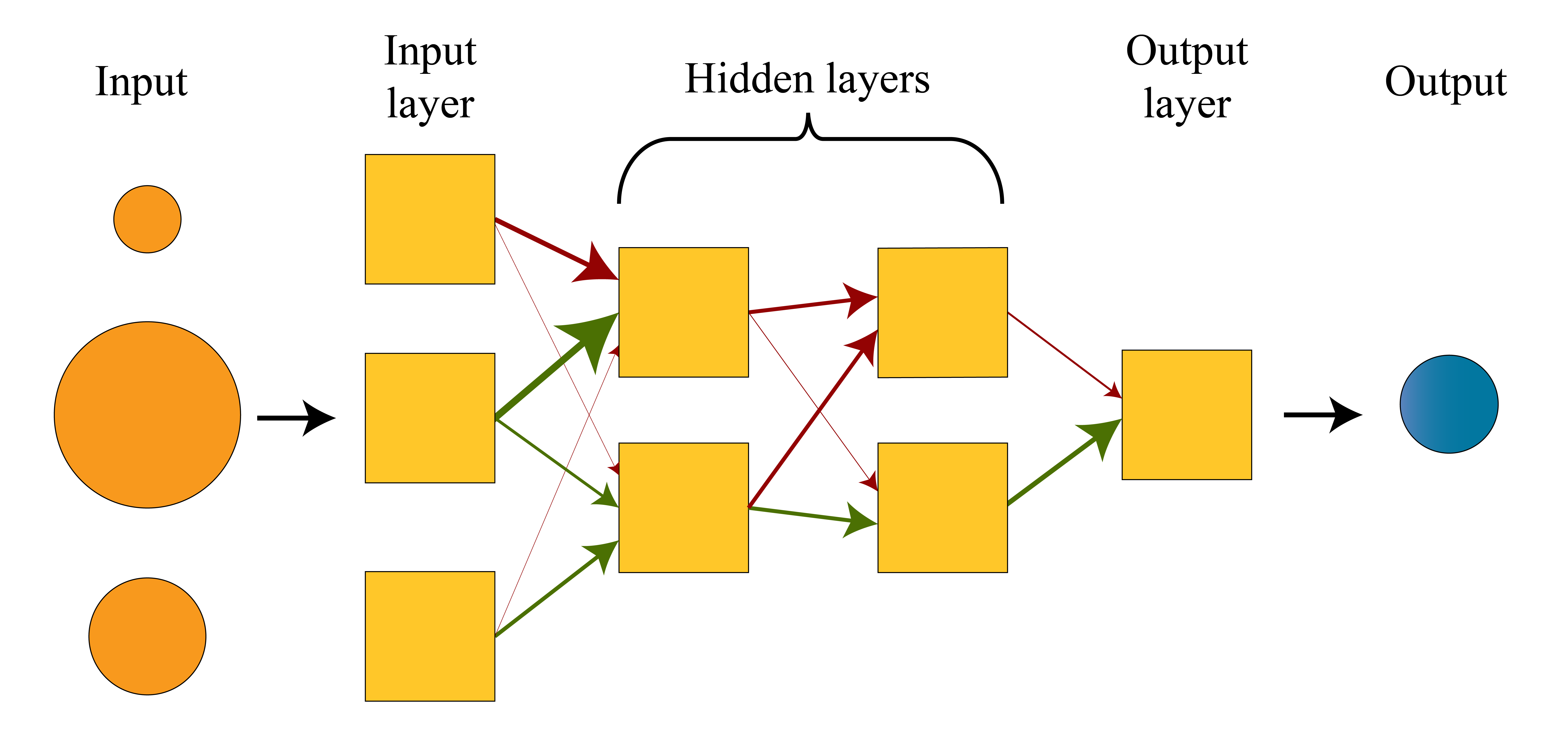}
  \caption{A Neural network consisting of an input layer, hidden layers and an output layer. The orange and blue circles represent input and output parameters respectively. The red and green arrows represent the weights that connect each layer in the neural network. }
  \label{fig:NN}
\end{figure}

\begin{figure*}
\centering
  \includegraphics[width=0.9\textwidth]{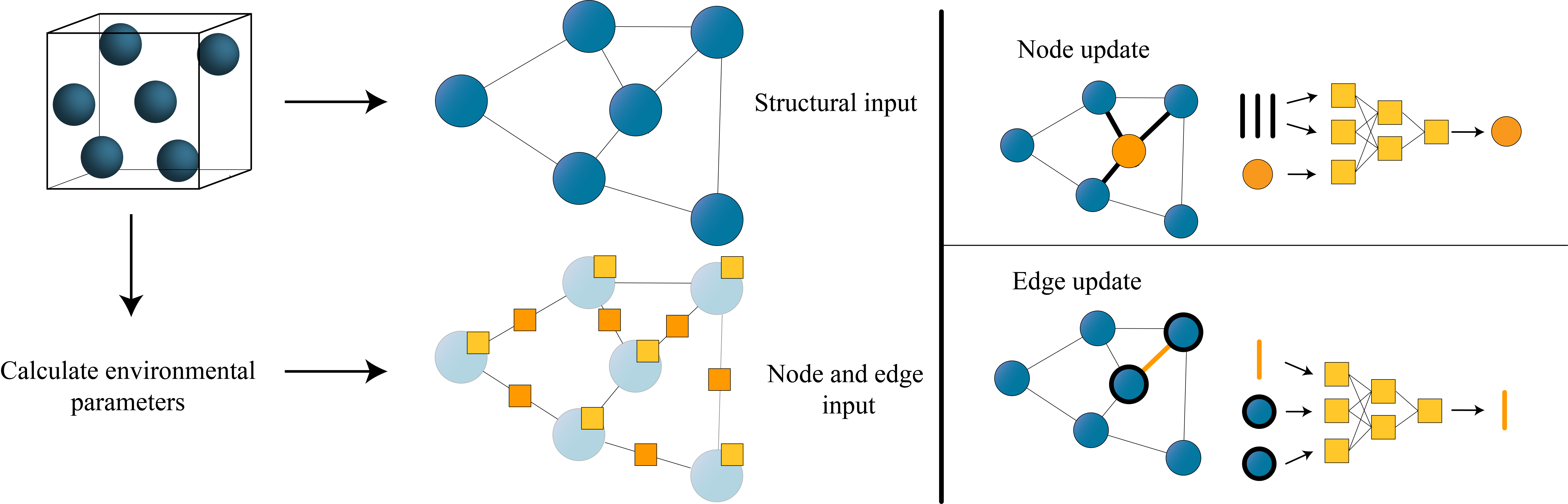}
  \caption{(left) Mapping between the initial configuration data and the graph structure. From the configuration, we determine nearest neighbors in order to set up a graph structure, and then calculate structural parameters to fill in the data at each node and edge. 
  (right)  Node and edge update in a graph layer. To update the node data, the added and averaged parameters of the connected edges are given to the node neural network, together with the information of the node itself. For the edge update, the information of both neighbouring nodes, together with the edge information itself is given to the edge neural network. The output of the node/edge network is an updated value of the node/edge.
  }
  \label{fig:Graphsetup}
\end{figure*}

Neural networks are loosely based on the biological neural networks that make up our brains. A neural network consists of multiple layers of connected nodes, see Fig. \ref{fig:NN}, which mimic the neurons and synapses in the brain. The first and last layer are respectively called the input and output layer, which in our case take the structural parameters of each particle as an input, and give the predicted propensity for a certain time as the output. All the layers that lie between the input and output layer are called hidden layers. In a fully connected feed-forward neural network, as we use here, each node in a specific layer is connected to all the nodes in the following layer. \\

Due to the connections between nodes, information can be passed through the neural network.   Each connection between nodes is associated with a so called `weight'. The values of the nodes in the hidden layer and the output layer are found by multiplying the values of the nodes in the previous layer with the associated weights and adding a bias. The result is then passed to a non-linear function, in our case a Rectified Linear Unit (ReLU) \cite{bookmachinelearning}, to yield the value of the node. Hence, the value of a node $a_m$ in layer \textit{l} is calculated as
\begin{equation}
a_m^{(l)} = f\left(\sum_n w^{(l)}_{mn}a_n^{(l-1)} +b^{(l)}_m\right).
\label{eq:weightedinput}
\end{equation}\\
Here $f$ is the ReLU function, $w^{(l)}_{mn}$ is the weight associated with the connection between nodes \textit{n} and \textit{m}, $a_n^{(l-1)}$ is the information of node \textit{n} in layer $l-1$ and $b^{(l)}_m$ is the bias associated with node $a_m^{(l)}$. The summation over \textit{n} goes over all nodes in layer $l-1$.

We train the neural networks using the Python package Pytorch\cite{PYTORCH}, and in particular use an Adam optimizer\cite{adam}. This optimizer is an extension to the stochastic gradient descent procedure, and is used to find an efficient path to a locally optimal set of weights and biases via backpropagation \cite{bookmachinelearning}. 

For neural networks there are many more hyperparameters that can be tuned, including the number of layers, and the number of nodes in each layer, as well as parameters associated with the learning process, such as the learning rate and the batch size. 
As the neural network turns out to be very sensitive to overfitting, we also run the neural network with ridge regression and drop-out \cite{dropout}.

\subsection{Graph neural networks}

GNNs \cite{GNN1,GNN2,GNN3} are a relatively new class of machine learning techniques that combine neural networks with a graph-like data structure. As there are many variations of GNNs, our description is necessarily somewhat specific to the GNN we use in this paper.  

In contrast to LR and NNs, which consider each particle as a separate training example, the GNN takes in an entire configuration at once. As a result, the GNN does not make a propensity prediction on a single-particle basis, but instead simultaneously makes a prediction for all particles (of a chosen species) in a configuration of the system. 

In a GNN, input data is first mapped to a data structure which consists of a graph that holds numerical data at its nodes and edges (see Figure \ref{fig:Graphsetup}). In our case, each node corresponds to a particle in the configuration for which we are trying to predict propensities. When two particles are closer to each other than a certain distance $r_c = 2.1 \sigma_L$, the corresponding nodes in the graph are connected by an edge. 

% In order for the GNN to be as close as possible to the other two methods, at each node 

As an input, every node holds the structural parameters of the corresponding particle as well as its species, and each edge holds information about the distance between the two connected particles. 
Like a neural network, a GNN corresponds to a non-linear function of its input parameters, that gets calculated in multiple consecutive layers, called graph layers. Each layer itself is a non-linear function, which takes as an input a graph structure containing information on all nodes and edges, and outputs a new graph structure with the same graph topology but updated numerical data at the nodes and edges. For each node and edge, the update that takes place within this graph layer takes into account not only the information already at that  node or edge, but also data from neighboring edges or nodes, as illustrated in Fig. \ref{fig:Graphsetup}. 
The internal functions that perform this update consist of fully connected feed-forward neural networks as described above. In addition to these graph layers, the full GNN incorporates an encoder layer, which maps the input node data to the data structure used in the graph layers, and a decoder, which makes a propensity prediction for each particle of a chosen species, based on the updated node data. Both of these layers are standard feed-forward neural networks. Note that analogous to the GNN in Ref. \onlinecite{BAPST}, our GNN additionally provides each graph layer with information about the graph data output by the encoder. In other words, a node update takes into account i) the node's current values, ii) the aggregated current values of all edges connected to the node, and iii) the node's values as they were just after the encoder layer.

The core benefit of the GNN is that the prediction for the propensity of a given particle can include information about the structure of multiple shells of neighboring particles.  The distance over which information is included can be controlled by the number of included graph layers. As such, the GNN inherently includes the feature of recursively considering the average local structure of shells of neighbors.

\section{Results}
In this section we first look at each method independently, and explore the influence of the different hyperparameter settings, and the inclusion of different generations of structural order parameters. In all cases, to compare the predictions to the measured propensity we use the Pearson correlation coefficient.  

\subsection{Linear regression}
\begin{figure}
\centering
\includegraphics[width=0.45\textwidth]{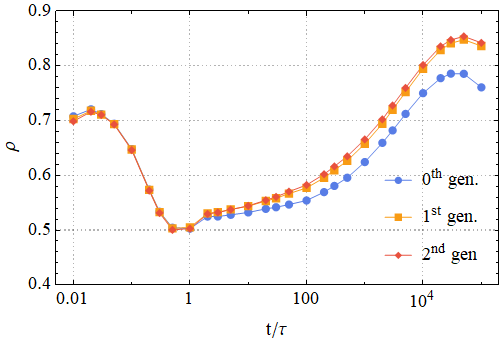}
\caption{Pearson's correlation coefficient $\rho$ between the dynamic propensity as predicted by linear regression and measured in simulations as a function of time, analyzed for three different generations of order parameters. Results are shown for the large particles.}
  \label{fig:pearsoncoefficentLR}
\end{figure}

The only hyperparameter for linear regression is the regularization parameter $\alpha$.  As optimizing the fit for a single parameter is trivial, here we only present results that correspond to the best choice of $\alpha$, optimized between $10^{-5}$ and $10^4$ for the large particles.   In Figure \ref{fig:pearsoncoefficentLR} we show
the linear regression  performance for different generations of order parameters. In all cases, when we refer to a generation, we include all lower generations as well. Note that these results are consistent with Ref.~\onlinecite{PaperEmanuele}.  We clearly see that the predictions from the zeroth generation of descriptors are significantly worse than the ones including higher-generation data, at least for longer times. In particular, we see that the information of the higher-order generations only starts to influence the performance when the system enters the caging regime. This is what we expected: before entering the caging regime not enough time has passed for particles to be influenced by particles from further away, meaning that higher-order generations will not add relevant information about the expected trajectories. 
Although adding the second generation of order parameters still improves the predictions for the propensity, the effect is small in comparison to the improvement of adding the first generation. Adding even higher generations (not shown here) does not significantly improve the performance beyond this point \cite{PaperEmanuele}.

\subsection{Neural networks}
\begin{figure}
\centering
\includegraphics[width=0.45\textwidth]{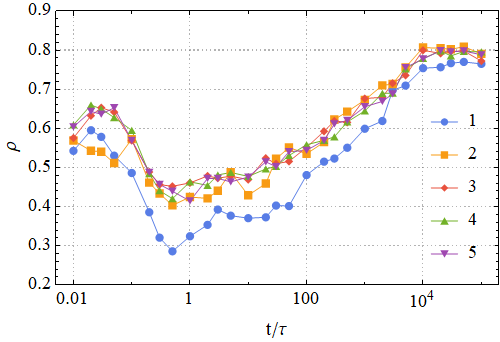}
 \caption{Correlation between the dynamic propensity as predicted by neural networks and measured in simulations as a function of time, analyzed for different hyperparameters as specified in Table \ref{tab:neuralnetworks}.  Results are shown for the large particles.}
  \label{fig:pearsoncoefficentNN}
\end{figure}

For the neural networks, in addition to the number of generations and regularization parameter, there are many hyperparameters to optimize.  
Since trying out all possible combination of settings would not be feasible, we limit ourselves to a small number of different combinations, shown in Table \ref{tab:neuralnetworks}. The network is trained on large and small particles separately in around 500 epochs.
In order to limit overfitting, after each epoch we evaluate the performance of the NN on the test data set, and eventually use the network that had the highest correlation for this test data. 
In total for each time interval and for each generation of structural parameters, we trained five networks.  Their performance is shown in Fig. \ref{fig:pearsoncoefficentNN}. Here, all networks are trained using three generations of structural parameters.

\begin{table}
\begin{center}
 \begin{tabular}{||c c c c c c||} 
 \hline
 Number& Batch size & Learning rate & Hidden layers& Drop-out & $\alpha$  \\ [0.5ex]
 \hline\hline
 1&50 & $10^{-4}$  & 1 (16) & 0 & 0 \\
 \hline
 2&50 &$10^{-4}$  & 3 (16,16,16)&0&0 \\ 
 \hline
  3&50 &$10^{-4}$  & 2 (16,16) &0&1.0\\ 
 \hline
  4&50 &$10^{-4}$  & 3 (16,16)&0.25&0 \\ 
 \hline
  5&50 &$10^{-4}$ & 3 (16,16)&0.25&0.01 \\ 
 \hline
\end{tabular}
\caption{Hyperparameters for different neural networks used to predict the propensity. Each row corresponds to a line in Fig. \ref{fig:pearsoncoefficentNN}. The \textit{Hidden layers} entry contains both information about the number of hidden layers in the network, as well as the number of nodes in each hidden layer (shown in brackets). \textit{Drop-out} shows the fraction of nodes in the second hidden layer that is set to zero during drop-out and $\alpha$ represents the value of the parameter associated with $L2-$regularization. }
 \label{tab:neuralnetworks}
\end{center}
\end{table}

Although the overall behavior of the neural network accuracy over time is similar to that found for linear regression, we see significant variation in performance between the different hyperparameter choices. In particular, we see that only using a single NN layer (blue line) or no regularization (blue and yellow lines) lead to worse performance, especially at shorter times. However, once we include at least two NN layers and regularization, the results are relatively robust: the other three lines essentially coincide except for noise. 

%The most probable explanation for this, is that the higher number of optimization parameters in the neural network inherently requires more training data. Indeed, by experimenting with different amounts of training data (not shown), we observe that more data leads to smoother predictions as a function of time.  This can also be observed by comparing the performance for small and large particles in Figure \ref{fig:pearsoncoefficentNN}: due to the abundance of small particles in the system, there is more training data for small particles, which results in smoother graphs.\\

%The sensitivity of the network to both the input parameters as well as the hyperparameters is a very clear expression of the sensitivity to overfitting. This statement is supported by the observation that the performance of the NN is mostly consistent between hyperparameter sets 3, 4, and 5. In the cases of all these performances we used some protection to overfitting, either in the form of regularization or drop-out. \\

\begin{figure}
\centering
\includegraphics[width=0.45\textwidth]{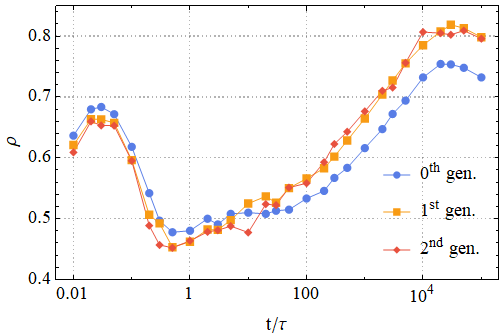}
  \caption{Correlation between the dynamic propensity as predicted by neural networks and measured in simulations as a function of time, analyzed for three different generations of order parameters.   Results are shown for the large particles.
  }
  \label{fig:pearsoncoefficentNN3gen}
\end{figure}

In order to examine the influence of different generations of order parameters on the performance, in Fig. \ref{fig:pearsoncoefficentNN3gen} we show the performance of the NN for different generations, with the hyperparameters optimized for each time interval. Contrary to what we saw in the case of linear regression, providing the NN with more generations of order parameters does not always improve the performance, something that is especially clear for short times. This is likely the result of the higher number of weights and biases that the NN training needs to optimize when higher-order descriptors are included.

\subsection{Graph neural networks}
%Moreover, due to the large number of weights and biases involved, GNNs are significantly more time-consuming to train and hence exploring a large number of hyperparameter optimisations is costly.
In Ref. \onlinecite{BAPST}, Bapst {\it et al.} demonstrated that hyperparameters did not play a large role in the accuracy of their GNNs for predicting dynamic propensity. Since the training of a GNN is considerably more expensive than a normal neural network, here we do not focus on optimizing the hyperparameters, but instead consider a few different variations of supplying data to the GNN. As a baseline, we consider a GNN with four graph layers, which considers the zero'th generation of structural parameters as the node inputs, and the absolute distance between particles as the edge inputs. The GNN is trained separately to predict the propensities of the large and small particles, but takes the information of all particles into account for both trainings. As a variation on this baseline, we also consider i) a GNN that predicts both the propensity of both the large and small species simultaneously, ii) a GNN that incorporates all three generations as node data, and iii) a GNN that uses the $x$, $y$, and $z$ components of the vectors between neighboring particles as edge data. A complete summary of all relevant (hyper)parameters is shown in Table \ref{tab:graphnetworks}.

%  The graphs are fed to the GNN in batches of 5, and we train for between 200 and 800 epochs. In order to check whether the GNN also benefits from higher generations of structural parameters, we train the network for both the first generation and the first three generations. 

\begin{figure}
\centering
\includegraphics[width=0.45\textwidth]{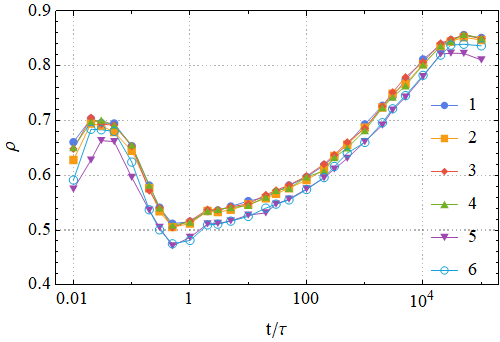}
\caption{Correlation between the dynamic propensity as predicted by graph neural networks and measured in simulations as a function of time, analyzed for different variations of the GNN model, which can be found in Table \ref{tab:graphnetworks}.}
  \label{fig:pearsoncoefficentGNN}
\end{figure}

In Fig. \ref{fig:pearsoncoefficentGNN}, we show the performance of the different variations of GNN.
Clearly, none of the changes made to the GNN inputs have a significant effect on the overall performance. The most significant difference is associated with the networks that trained both small and large particles simultaneously -- this adaptation performed worse for our system. In contrast to linear regression and neural networks, where taking into account additional generations clearly led to some improvement in the performance, evidently the inherent local averaging of the GNN is already sufficient to incorporate this type of structural information. Additionally, we observe that compared to the NN, the GNN is less sensitive to overfitting, resulting in relatively smooth lines. This is actually quite remarkable, since the number of weights and biases that need to be optimized in a GNN is significantly larger than in the case of a NN. Moreover, none of the training runs in Fig. \ref{fig:pearsoncoefficentGNN} used any regularization methods. The fact that GNNs, compared to NNs, have less trouble converging and are less sensitive to overfitting, might be due to the fact that they are trained and evaluated on entire snapshots at once. Realistically, in a snapshot we know that there are strong correlations in mobility between neighboring particles. The fact that the GNN can take into account the mobility of neighbouring particles, will likely lead to smoother variation of the predicted propensity in space than the NN or LR, which results in fewer outliers.

\begin{table*}
\setlength{\tabcolsep}{3.5pt}
 \hspace{-3cm}
\begin{center}
 \addtolength{\leftskip} {-2cm} % increase (absolute) value if needed
    \addtolength{\rightskip}{-2cm}
 \begin{tabular}{||c c c c c c c c c c c||} 
 \hline
Nr. &\makecell{Batch\\size} & \makecell{Learning\\rate}  & GL & \makecell{Node \\ enc H.L.} & \makecell{Edge\\ enc H.L. }  & \makecell{Node\\ H.L} & \makecell{Edge\\ H.L.} &\makecell{Node\\ gen}& \makecell{Edge\\ par} &\makecell{Tog or\\ Sep}\\ [0.5ex] 
 \hline\hline
  1&5 & $10^{-4}$  &4 & 2 (50, 50) & 1 (5) & \makecell{2 (30, 16)\\ 2 (16, 16)} &\makecell{2 (16, 16)\\ 2 (16, 16)} & 1  & \textit{r} & Sep  \\
  \hline
   2&5 & $10^{-4}$  &4 & 2 (50, 50) & 1 (5) & \makecell{2 (30, 16)\\ 2 (16, 16)}&\makecell{2 (16, 16)\\ 2 (16, 16)} & 3  & \textit{r}& Sep  \\
        \hline
      3&5 & $10^{-4}$  &4 & 2 (50, 50) & 1 (5) & \makecell{2 (30, 16)\\ 2 (16, 16)}&\makecell{2 (16, 16)\\ 2 (16, 16)} & 1  & \textit{x, y, z}& Sep  \\
   \hline
     4&5 & $10^{-3}$  &4 & 2 (50, 50) & 1 (5) & \makecell{2 (30, 16)\\ 2 (16, 16)}&\makecell{2 (16, 16)\\ 2 (16, 16)} & 1  & \textit{x, y, z}& Sep\\
 \hline
   5& 5 & $10^{-4}$  & 4 & 2 (50, 50) & 1 (5) & \makecell{2 (30, 16)\\ 2 (16, 16)}&\makecell{2 (16, 16)\\ 2 (16, 16)} & 1  & \textit{r}& Tog  \\
   \hline
  6& 3 & $10^{-4}$  & 3 & 2 (50, 50) & 1 (5) & \makecell{2 (30, 16)\\ 2 (16, 16)}&\makecell{2 (16, 16)\\ 2 (16, 16)} & 1  & \textit{r}& Tog  \\
  \hline

\end{tabular}
   \caption{Hyperparameters for different graph neural networks used to predict the propensity. Each row corresponds to a line in Fig. \ref{fig:pearsoncoefficentGNN}. In the table the following abbreviations are used: \textit{GL} is the number of graph layers,\textit{ Node enc H.L.} and\textit{ Edge enc H.L.} represent respectively the number of hidden layers in the node and edge encoder (for both, the number of input parameters is equal to the number of parameters associated with a node or an edge, while the number of output parameters is equal to 10.\textit{ Node H.L} and\textit{ Edge H.L}. show the number of hidden layers for the respectively the node and edge hidden layers together with the number of nodes in each layers (the number of output nodes for each of these networks is equal to 10), \textit{Node gen} represents up to how many generations of structural order parameters we provide the GNN with, \textit{Edge par} indicates whether the edge input is given by the absolute distance between particles (\textit{r}), or the vector distance (\textit{x}, \textit{y}, \textit{z}). Finally \textit{Tog or Sep} represents whether we train the network together (Tog) for small and large particles, or train two separate networks (Sep). }
\label{tab:graphnetworks}
\end{center}
\end{table*}

\subsection{Comparing the three methods}

\begin{figure}
\centering
\begin{tabular}{lc}
a)&\\
&\includegraphics[width=0.45\textwidth]{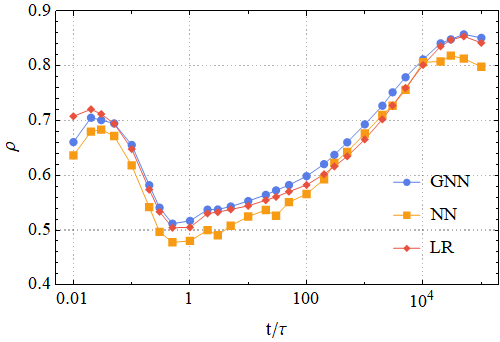}\\
b)&\\
&\includegraphics[width=0.45\textwidth]{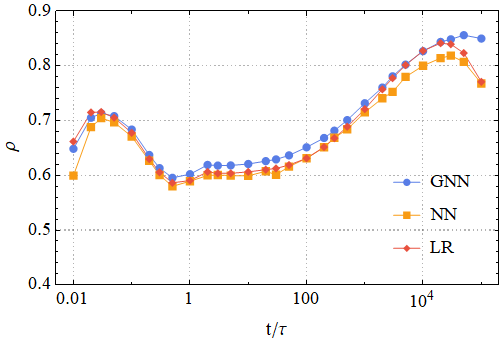}
\end{tabular}
  \caption{Comparison of the accuracy of the dynamic propensity prediction over time as obtained with LR, NN and GNN. Results are shown for both the (a) large and (b) small particles. For each point in time we use for each method the set of hyperparameters that resulted in the best performance.}
  \label{fig:pearsoncoefficenttotal}
\end{figure}

Finally, in Fig. \ref{fig:pearsoncoefficenttotal}, we compare the performance of all three methods. Again, we take for each time interval the best-performing result from either LR, NN, or GNN. The results of the three methods are remarkably similar, suggesting that all three approaches are capable of extracting essentially the same information from the input data. Overall, the NN approach performs the worst, likely due to an inability to find the globally optimal solution to its training problem. Intuitively, the fit from linear regression could be reproduced essentially exactly by the NN, with a sufficiently good optimization. Hence, at any point where the NN performs less well than the LR solution, there is at least some failure to fully optimize the network.

As we saw earlier, GNNs are less sensitive to their hyperparameters than NNs, and converge more easily. Moreover, during intermediate times in the caging regime and the beginning of the diffusive regime GNNs slightly outperform LR. This implies that the averaging that takes place in a GNN provides the network with slightly different information than the averaged parameters of the first and second generation. However, the improvement is extremely limited, and comes at the cost of a significantly more computationally costly training process. To illustrate this: for one choice of hyperparameters and the full set of time intervals, a typical training process takes approximately 3 minutes for LR, 24 hours for NNs, and 6.5 hours for GNNs. Note that these times are achieved by a laptop CPU for the LR, while the NN and GNN trainings made use of an Nvidia GeForce RTX 2080Ti GPU.
We conclude that, given the discussion above, linear regression is the preferred method; it is fast, robust, and provides accurate predictions.

\section{Conclusion}

In this paper, we compared three different ML methods for predicting the dynamic propensity of a glassy system at different times, namely linear regression, neural networks, and graph neural networks. We find surprisingly little difference in their performance over the full range of time intervals considered. 
The intuitive conclusion one can draw from the similar results of the three methods is that our prediction, at this point, is limited not by our fitting approach, but rather by the information contained in the set of structural order parameters. On the bright side, this means that given this set of descriptors, a simple and efficient ML method like linear regression is sufficient for essentially optimal predictions. On the other hand, it suggests that more advanced ML techniques are not likely to provide a solution to the question of how to further improve the prediction of dynamics in these systems. 
The main question that remains is: what information are we currently missing in order to improve our ability to predict the dynamic propensity, in particular in the caging regime, where the correlation between prediction and reality is currently minimal? While answering this question will require further research, our results here suggest that linear regression is likely a sufficient method for evaluating the predictive capabilities of new sets of structural order parameters.

\section*{References}

\bibliography{myref}

\begin{thebibliography}{23}%
\makeatletter
\providecommand \@ifxundefined [1]{%
 \@ifx{#1\undefined}
}%
\providecommand \@ifnum [1]{%
 \ifnum #1\expandafter \@firstoftwo
 \else \expandafter \@secondoftwo
 \fi
}%
\providecommand \@ifx [1]{%
 \ifx #1\expandafter \@firstoftwo
 \else \expandafter \@secondoftwo
 \fi
}%
\providecommand \natexlab [1]{#1}%
\providecommand \enquote  [1]{``#1''}%
\providecommand \bibnamefont  [1]{#1}%
\providecommand \bibfnamefont [1]{#1}%
\providecommand \citenamefont [1]{#1}%
\providecommand \href@noop [0]{\@secondoftwo}%
\providecommand \href [0]{\begingroup \@sanitize@url \@href}%
\providecommand \@href[1]{\@@startlink{#1}\@@href}%
\providecommand \@@href[1]{\endgroup#1\@@endlink}%
\providecommand \@sanitize@url [0]{\catcode `\\12\catcode `\$12\catcode
  `\&12\catcode `\#12\catcode `\^12\catcode `\_12\catcode `\%12\relax}%
\providecommand \@@startlink[1]{}%
\providecommand \@@endlink[0]{}%
\providecommand \url  [0]{\begingroup\@sanitize@url \@url }%
\providecommand \@url [1]{\endgroup\@href {#1}{\urlprefix }}%
\providecommand \urlprefix  [0]{URL }%
\providecommand \Eprint [0]{\href }%
\providecommand \doibase [0]{http://dx.doi.org/}%
\providecommand \selectlanguage [0]{\@gobble}%
\providecommand \bibinfo  [0]{\@secondoftwo}%
\providecommand \bibfield  [0]{\@secondoftwo}%
\providecommand \translation [1]{[#1]}%
\providecommand \BibitemOpen [0]{}%
\providecommand \bibitemStop [0]{}%
\providecommand \bibitemNoStop [0]{.\EOS\space}%
\providecommand \EOS [0]{\spacefactor3000\relax}%
\providecommand \BibitemShut  [1]{\csname bibitem#1\endcsname}%
\let\auto@bib@innerbib\@empty
%</preamble>
\bibitem [{\citenamefont {Berthier}\ and\ \citenamefont
  {Biroli}(2011)}]{theoryglasses}%
  \BibitemOpen
  \bibfield  {author} {\bibinfo {author} {\bibfnamefont {L.}~\bibnamefont
  {Berthier}}\ and\ \bibinfo {author} {\bibfnamefont {G.}~\bibnamefont
  {Biroli}},\ }\bibfield  {title} {\enquote {\bibinfo {title} {Theoretical
  perspective on the glass transition and amorphous materials},}\ }\href@noop
  {} {\bibfield  {journal} {\bibinfo  {journal} {Reviews of Modern Physics}\
  }\textbf {\bibinfo {volume} {83}},\ \bibinfo {pages} {587} (\bibinfo {year}
  {2011})}\BibitemShut {NoStop}%
\bibitem [{\citenamefont {Royall}\ and\ \citenamefont
  {Williams}(2015)}]{patty}%
  \BibitemOpen
  \bibfield  {author} {\bibinfo {author} {\bibfnamefont {C.}~\bibnamefont
  {Royall}}\ and\ \bibinfo {author} {\bibfnamefont {S.~R.}\ \bibnamefont
  {Williams}},\ }\bibfield  {title} {\enquote {\bibinfo {title} {The role of
  local structure in dynamical arrest},}\ }\href@noop {} {\bibfield  {journal}
  {\bibinfo  {journal} {Physics Reports}\ }\textbf {\bibinfo {volume} {560}},\
  \bibinfo {pages} {1} (\bibinfo {year} {2015})}\BibitemShut {NoStop}%
\bibitem [{\citenamefont {Tanaka}\ \emph {et~al.}(2019)\citenamefont {Tanaka},
  \citenamefont {Tong}, \citenamefont {Shi},\ and\ \citenamefont
  {Russo}}]{tanaka2019revealing}%
  \BibitemOpen
  \bibfield  {author} {\bibinfo {author} {\bibfnamefont {H.}~\bibnamefont
  {Tanaka}}, \bibinfo {author} {\bibfnamefont {H.}~\bibnamefont {Tong}},
  \bibinfo {author} {\bibfnamefont {R.}~\bibnamefont {Shi}}, \ and\ \bibinfo
  {author} {\bibfnamefont {J.}~\bibnamefont {Russo}},\ }\bibfield  {title}
  {\enquote {\bibinfo {title} {Revealing key structural features hidden in
  liquids and glasses},}\ }\href@noop {} {\bibfield  {journal} {\bibinfo
  {journal} {Nature Reviews Physics}\ }\textbf {\bibinfo {volume} {1}},\
  \bibinfo {pages} {333--348} (\bibinfo {year} {2019})}\BibitemShut {NoStop}%
\bibitem [{\citenamefont {Cubuk}\ \emph {et~al.}(2015)\citenamefont {Cubuk},
  \citenamefont {S.S.Schoenholz}, \citenamefont {Rieser}, \citenamefont
  {Malone}, \citenamefont {Rottler}, \citenamefont {Durian}, \citenamefont
  {Kaxiras},\ and\ \citenamefont {Liu}}]{structurevsdynamics1}%
  \BibitemOpen
  \bibfield  {author} {\bibinfo {author} {\bibfnamefont {E.~D.}\ \bibnamefont
  {Cubuk}}, \bibinfo {author} {\bibnamefont {S.S.Schoenholz}}, \bibinfo
  {author} {\bibfnamefont {J.~M.}\ \bibnamefont {Rieser}}, \bibinfo {author}
  {\bibfnamefont {B.~D.}\ \bibnamefont {Malone}}, \bibinfo {author}
  {\bibfnamefont {J.}~\bibnamefont {Rottler}}, \bibinfo {author} {\bibfnamefont
  {D.~J.}\ \bibnamefont {Durian}}, \bibinfo {author} {\bibfnamefont
  {E.}~\bibnamefont {Kaxiras}}, \ and\ \bibinfo {author} {\bibfnamefont
  {A.~J.}\ \bibnamefont {Liu}},\ }\bibfield  {title} {\enquote {\bibinfo
  {title} {Identifying structural flow defects in disordered solids using
  machine-learning methods},}\ }\href@noop {} {\bibfield  {journal} {\bibinfo
  {journal} {Physical Review Letters}\ }\textbf {\bibinfo {volume} {114}},\
  \bibinfo {pages} {108001} (\bibinfo {year} {2015})}\BibitemShut {NoStop}%
\bibitem [{\citenamefont {Bapst}\ \emph {et~al.}(2020)\citenamefont {Bapst},
  \citenamefont {Keck}, \citenamefont {Grabska-Barwi{\'n}ska}, \citenamefont
  {Donner}, \citenamefont {Cubuk}, \citenamefont {Schoenholz}, \citenamefont
  {Obika}, \citenamefont {Nelson}, \citenamefont {Back}, \citenamefont
  {Hassabis},\ and\ \citenamefont {Kohli}}]{BAPST}%
  \BibitemOpen
  \bibfield  {author} {\bibinfo {author} {\bibfnamefont {V.}~\bibnamefont
  {Bapst}}, \bibinfo {author} {\bibfnamefont {T.}~\bibnamefont {Keck}},
  \bibinfo {author} {\bibfnamefont {A.}~\bibnamefont {Grabska-Barwi{\'n}ska}},
  \bibinfo {author} {\bibfnamefont {C.}~\bibnamefont {Donner}}, \bibinfo
  {author} {\bibfnamefont {E.~D.}\ \bibnamefont {Cubuk}}, \bibinfo {author}
  {\bibfnamefont {S.~S.}\ \bibnamefont {Schoenholz}}, \bibinfo {author}
  {\bibfnamefont {A.}~\bibnamefont {Obika}}, \bibinfo {author} {\bibfnamefont
  {A.~W.~R.}\ \bibnamefont {Nelson}}, \bibinfo {author} {\bibfnamefont
  {T.}~\bibnamefont {Back}}, \bibinfo {author} {\bibfnamefont {D.}~\bibnamefont
  {Hassabis}}, \ and\ \bibinfo {author} {\bibfnamefont {P.}~\bibnamefont
  {Kohli}},\ }\bibfield  {title} {\enquote {\bibinfo {title} {Unveiling the
  predictive power of static structure in glassy systems},}\ }\href@noop {}
  {\bibfield  {journal} {\bibinfo  {journal} {Nature Physics}\ }\textbf
  {\bibinfo {volume} {16}},\ \bibinfo {pages} {448} (\bibinfo {year}
  {2020})}\BibitemShut {NoStop}%
\bibitem [{\citenamefont {Boattini}\ \emph {et~al.}(2020)\citenamefont
  {Boattini}, \citenamefont {Mar{\'i}n-Aguilar}, \citenamefont {Mitra},
  \citenamefont {Foffi}, \citenamefont {Smallenburg},\ and\ \citenamefont
  {Filion}}]{glassydnamics2}%
  \BibitemOpen
  \bibfield  {author} {\bibinfo {author} {\bibfnamefont {E.}~\bibnamefont
  {Boattini}}, \bibinfo {author} {\bibfnamefont {S.}~\bibnamefont
  {Mar{\'i}n-Aguilar}}, \bibinfo {author} {\bibfnamefont {S.}~\bibnamefont
  {Mitra}}, \bibinfo {author} {\bibfnamefont {G.}~\bibnamefont {Foffi}},
  \bibinfo {author} {\bibfnamefont {F.}~\bibnamefont {Smallenburg}}, \ and\
  \bibinfo {author} {\bibfnamefont {L.}~\bibnamefont {Filion}},\ }\bibfield
  {title} {\enquote {\bibinfo {title} {Autonomously revealing hidden local
  structures in supercooled liquids},}\ }\href@noop {} {\bibfield  {journal}
  {\bibinfo  {journal} {Nature Communications}\ }\textbf {\bibinfo {volume}
  {11}},\ \bibinfo {pages} {5479} (\bibinfo {year} {2020})}\BibitemShut
  {NoStop}%
\bibitem [{\citenamefont {Paret}, \citenamefont {Jack},\ and\ \citenamefont
  {Coslovich}(2020)}]{structurevsdynamics5}%
  \BibitemOpen
  \bibfield  {author} {\bibinfo {author} {\bibfnamefont {J.}~\bibnamefont
  {Paret}}, \bibinfo {author} {\bibfnamefont {R.~L.}\ \bibnamefont {Jack}}, \
  and\ \bibinfo {author} {\bibfnamefont {D.}~\bibnamefont {Coslovich}},\
  }\bibfield  {title} {\enquote {\bibinfo {title} {Assessing the structural
  heterogeneity of supercooled liquids through community inference},}\
  }\href@noop {} {\bibfield  {journal} {\bibinfo  {journal} {The Journal of
  Chemical Physics}\ }\textbf {\bibinfo {volume} {152}},\ \bibinfo {pages}
  {144502} (\bibinfo {year} {2020})}\BibitemShut {NoStop}%
\bibitem [{\citenamefont {Boattini}, \citenamefont {Smallenburg},\ and\
  \citenamefont {Filion}(2021)}]{PaperEmanuele}%
  \BibitemOpen
  \bibfield  {author} {\bibinfo {author} {\bibfnamefont {E.}~\bibnamefont
  {Boattini}}, \bibinfo {author} {\bibfnamefont {F.}~\bibnamefont
  {Smallenburg}}, \ and\ \bibinfo {author} {\bibfnamefont {L.}~\bibnamefont
  {Filion}},\ }\bibfield  {title} {\enquote {\bibinfo {title} {Averaging local
  structure to predict the dynamic propensity in supercooled liquids},}\
  }\href@noop {} {\bibfield  {journal} {\bibinfo  {journal} {Physical Review
  Letters}\ }\textbf {\bibinfo {volume} {127}},\ \bibinfo {pages} {088007}
  (\bibinfo {year} {2021})}\BibitemShut {NoStop}%
\bibitem [{\citenamefont {Schoenholz}\ \emph {et~al.}(2016)\citenamefont
  {Schoenholz}, \citenamefont {Cubuk}, \citenamefont {Sussman}, \citenamefont
  {Kaxiras},\ and\ \citenamefont {Liu}}]{andrea}%
  \BibitemOpen
  \bibfield  {author} {\bibinfo {author} {\bibfnamefont {S.~S.}\ \bibnamefont
  {Schoenholz}}, \bibinfo {author} {\bibfnamefont {E.~D.}\ \bibnamefont
  {Cubuk}}, \bibinfo {author} {\bibfnamefont {D.~M.}\ \bibnamefont {Sussman}},
  \bibinfo {author} {\bibfnamefont {E.}~\bibnamefont {Kaxiras}}, \ and\
  \bibinfo {author} {\bibfnamefont {A.~J.}\ \bibnamefont {Liu}},\ }\bibfield
  {title} {\enquote {\bibinfo {title} {{A structural approach to relaxation in
  glassy liquids}},}\ }\href@noop {} {\bibfield  {journal} {\bibinfo  {journal}
  {Nature Physics}\ }\textbf {\bibinfo {volume} {12}},\ \bibinfo {pages} {469}
  (\bibinfo {year} {2016})}\BibitemShut {NoStop}%
\bibitem [{\citenamefont {Richard}\ \emph {et~al.}(2020)\citenamefont
  {Richard}, \citenamefont {Ozawa}, \citenamefont {Patinet}, \citenamefont
  {Stanifer}, \citenamefont {Shang}, \citenamefont {Ridout}, \citenamefont
  {Xu}, \citenamefont {Zhang}, \citenamefont {Morse}, \citenamefont {Barrat}
  \emph {et~al.}}]{richard2020predicting}%
  \BibitemOpen
  \bibfield  {author} {\bibinfo {author} {\bibfnamefont {D.}~\bibnamefont
  {Richard}}, \bibinfo {author} {\bibfnamefont {M.}~\bibnamefont {Ozawa}},
  \bibinfo {author} {\bibfnamefont {S.}~\bibnamefont {Patinet}}, \bibinfo
  {author} {\bibfnamefont {E.}~\bibnamefont {Stanifer}}, \bibinfo {author}
  {\bibfnamefont {B.}~\bibnamefont {Shang}}, \bibinfo {author} {\bibfnamefont
  {S.}~\bibnamefont {Ridout}}, \bibinfo {author} {\bibfnamefont
  {B.}~\bibnamefont {Xu}}, \bibinfo {author} {\bibfnamefont {G.}~\bibnamefont
  {Zhang}}, \bibinfo {author} {\bibfnamefont {P.}~\bibnamefont {Morse}},
  \bibinfo {author} {\bibfnamefont {J.-L.}\ \bibnamefont {Barrat}},  \emph
  {et~al.},\ }\bibfield  {title} {\enquote {\bibinfo {title} {Predicting
  plasticity in disordered solids from structural indicators},}\ }\href@noop {}
  {\bibfield  {journal} {\bibinfo  {journal} {Physical Review Materials}\
  }\textbf {\bibinfo {volume} {4}},\ \bibinfo {pages} {113609} (\bibinfo {year}
  {2020})}\BibitemShut {NoStop}%
\bibitem [{\citenamefont {Mar{\'i}n-Aguilar}\ \emph {et~al.}(2020)\citenamefont
  {Mar{\'i}n-Aguilar}, \citenamefont {Wensink}, \citenamefont {Foffi},\ and\
  \citenamefont {Smallenburg}}]{Tetrahedrality}%
  \BibitemOpen
  \bibfield  {author} {\bibinfo {author} {\bibfnamefont {S.}~\bibnamefont
  {Mar{\'i}n-Aguilar}}, \bibinfo {author} {\bibfnamefont {H.~H.}\ \bibnamefont
  {Wensink}}, \bibinfo {author} {\bibfnamefont {G.}~\bibnamefont {Foffi}}, \
  and\ \bibinfo {author} {\bibfnamefont {F.}~\bibnamefont {Smallenburg}},\
  }\bibfield  {title} {\enquote {\bibinfo {title} {Tetrahedrality dictates
  dynamics in hard sphere mixtures},}\ }\href@noop {} {\bibfield  {journal}
  {\bibinfo  {journal} {Physical Review Letters}\ }\textbf {\bibinfo {volume}
  {124}},\ \bibinfo {pages} {208005} (\bibinfo {year} {2020})}\BibitemShut
  {NoStop}%
\bibitem [{\citenamefont {Rapaport}(2009)}]{EDMD}%
  \BibitemOpen
  \bibfield  {author} {\bibinfo {author} {\bibfnamefont {D.~C.}\ \bibnamefont
  {Rapaport}},\ }\bibfield  {title} {\enquote {\bibinfo {title} {{The
  Event-Driven Approach to {N}-Body Simulation}},}\ }\href@noop {} {\bibfield
  {journal} {\bibinfo  {journal} {Progress of Theoretical Physics Supplement}\
  }\textbf {\bibinfo {volume} {178}},\ \bibinfo {pages} {5--14} (\bibinfo
  {year} {2009})}\BibitemShut {NoStop}%
\bibitem [{\citenamefont {Widmer-Cooper}, \citenamefont {Harrowell},\ and\
  \citenamefont {Fynewever}(2004)}]{dynamicpropensity}%
  \BibitemOpen
  \bibfield  {author} {\bibinfo {author} {\bibfnamefont {A.}~\bibnamefont
  {Widmer-Cooper}}, \bibinfo {author} {\bibfnamefont {P.}~\bibnamefont
  {Harrowell}}, \ and\ \bibinfo {author} {\bibfnamefont {H.}~\bibnamefont
  {Fynewever}},\ }\bibfield  {title} {\enquote {\bibinfo {title} {How
  reproducible are dynamic heterogeneities in a supercooled liquid?}}\
  }\href@noop {} {\bibfield  {journal} {\bibinfo  {journal} {Physical Review
  Letters}\ }\textbf {\bibinfo {volume} {93}},\ \bibinfo {pages} {135701}
  (\bibinfo {year} {2004})}\BibitemShut {NoStop}%
\bibitem [{\citenamefont {Widmer-Cooper}\ and\ \citenamefont
  {Harrowell}(2007)}]{propensity1}%
  \BibitemOpen
  \bibfield  {author} {\bibinfo {author} {\bibfnamefont {A.}~\bibnamefont
  {Widmer-Cooper}}\ and\ \bibinfo {author} {\bibfnamefont {P.}~\bibnamefont
  {Harrowell}},\ }\bibfield  {title} {\enquote {\bibinfo {title} {On the study
  of collective dynamics in supercooled liquids through the statistics of the
  isoconfigurational ensemble},}\ }\href@noop {} {\bibfield  {journal}
  {\bibinfo  {journal} {The Journal of Chemical Physics}\ }\textbf {\bibinfo
  {volume} {126}},\ \bibinfo {pages} {154503} (\bibinfo {year}
  {2007})}\BibitemShut {NoStop}%
\bibitem [{\citenamefont {Lechner}\ and\ \citenamefont {Dellago}(2008)}]{BOPS}%
  \BibitemOpen
  \bibfield  {author} {\bibinfo {author} {\bibfnamefont {W.}~\bibnamefont
  {Lechner}}\ and\ \bibinfo {author} {\bibfnamefont {C.}~\bibnamefont
  {Dellago}},\ }\bibfield  {title} {\enquote {\bibinfo {title} {Accurate
  determination of crystal structures based on averaged local bond order
  parameters},}\ }\href@noop {} {\bibfield  {journal} {\bibinfo  {journal} {The
  Journal of Chemical Physics}\ }\textbf {\bibinfo {volume} {129}},\ \bibinfo
  {pages} {114707} (\bibinfo {year} {2008})}\BibitemShut {NoStop}%
\bibitem [{\citenamefont {.J.Steinhardt}, \citenamefont {Nelson},\ and\
  \citenamefont {Ronchetti}(1983)}]{BOPSold}%
  \BibitemOpen
  \bibfield  {author} {\bibinfo {author} {\bibfnamefont {P.}~\bibnamefont
  {.J.Steinhardt}}, \bibinfo {author} {\bibfnamefont {D.~R.}\ \bibnamefont
  {Nelson}}, \ and\ \bibinfo {author} {\bibfnamefont {M.}~\bibnamefont
  {Ronchetti}},\ }\bibfield  {title} {\enquote {\bibinfo {title}
  {Bond-orientational order in liquids and glasses},}\ }\href@noop {}
  {\bibfield  {journal} {\bibinfo  {journal} {Physical Review B}\ }\textbf
  {\bibinfo {volume} {28}},\ \bibinfo {pages} {784} (\bibinfo {year}
  {1983})}\BibitemShut {NoStop}%
\bibitem [{\citenamefont {Bishop}(2006)}]{bookmachinelearning}%
  \BibitemOpen
  \bibfield  {author} {\bibinfo {author} {\bibfnamefont {C.~M.}\ \bibnamefont
  {Bishop}},\ }\href@noop {} {\emph {\bibinfo {title} {Pattern Recognition and
  Machine Learning (Information Science and Statistics)}}}\ (\bibinfo
  {publisher} {Springer-Verlag},\ \bibinfo {year} {2006})\BibitemShut {NoStop}%
\bibitem [{\citenamefont {Paszke}\ \emph {et~al.}(2019)\citenamefont {Paszke},
  \citenamefont {Gross}, \citenamefont {Massa}, \citenamefont {Lerer},
  \citenamefont {Bradbury}, \citenamefont {Chanan}, \citenamefont {Killeen},
  \citenamefont {Lin}, \citenamefont {Gimelshein}, \citenamefont {Antiga} \emph
  {et~al.}}]{PYTORCH}%
  \BibitemOpen
  \bibfield  {author} {\bibinfo {author} {\bibfnamefont {A.}~\bibnamefont
  {Paszke}}, \bibinfo {author} {\bibfnamefont {S.}~\bibnamefont {Gross}},
  \bibinfo {author} {\bibfnamefont {F.}~\bibnamefont {Massa}}, \bibinfo
  {author} {\bibfnamefont {A.}~\bibnamefont {Lerer}}, \bibinfo {author}
  {\bibfnamefont {J.}~\bibnamefont {Bradbury}}, \bibinfo {author}
  {\bibfnamefont {G.}~\bibnamefont {Chanan}}, \bibinfo {author} {\bibfnamefont
  {T.}~\bibnamefont {Killeen}}, \bibinfo {author} {\bibfnamefont
  {Z.}~\bibnamefont {Lin}}, \bibinfo {author} {\bibfnamefont {N.}~\bibnamefont
  {Gimelshein}}, \bibinfo {author} {\bibfnamefont {L.}~\bibnamefont {Antiga}},
  \emph {et~al.},\ }\bibfield  {title} {\enquote {\bibinfo {title} {Pytorch: An
  imperative style, high-performance deep learning library},}\ }\href@noop {}
  {\bibfield  {journal} {\bibinfo  {journal} {Advances in neural information
  processing systems}\ }\textbf {\bibinfo {volume} {32}},\ \bibinfo {pages}
  {8026} (\bibinfo {year} {2019})}\BibitemShut {NoStop}%
\bibitem [{\citenamefont {Kingma}\ and\ \citenamefont {Ba}(2014)}]{adam}%
  \BibitemOpen
  \bibfield  {author} {\bibinfo {author} {\bibfnamefont {D.~P.}\ \bibnamefont
  {Kingma}}\ and\ \bibinfo {author} {\bibfnamefont {J.}~\bibnamefont {Ba}},\
  }\bibfield  {title} {\enquote {\bibinfo {title} {Adam: A method for
  stochastic optimization},}\ }\href@noop {} {\bibfield  {journal} {\bibinfo
  {journal} {arXiv preprint arXiv:1412.6980}\ } (\bibinfo {year}
  {2014})}\BibitemShut {NoStop}%
\bibitem [{\citenamefont {Srivastava}\ \emph {et~al.}(2014)\citenamefont
  {Srivastava}, \citenamefont {Hinton}, \citenamefont {Krizhevsky},
  \citenamefont {Sutskever},\ and\ \citenamefont {Salakhutdinov}}]{dropout}%
  \BibitemOpen
  \bibfield  {author} {\bibinfo {author} {\bibfnamefont {N.}~\bibnamefont
  {Srivastava}}, \bibinfo {author} {\bibfnamefont {G.}~\bibnamefont {Hinton}},
  \bibinfo {author} {\bibfnamefont {A.}~\bibnamefont {Krizhevsky}}, \bibinfo
  {author} {\bibfnamefont {I.}~\bibnamefont {Sutskever}}, \ and\ \bibinfo
  {author} {\bibfnamefont {R.}~\bibnamefont {Salakhutdinov}},\ }\bibfield
  {title} {\enquote {\bibinfo {title} {Dropout: A simple way to prevent neural
  networks from overfitting},}\ }\href@noop {} {\bibfield  {journal} {\bibinfo
  {journal} {Journal of Machine Learning Research}\ }\textbf {\bibinfo {volume}
  {15}},\ \bibinfo {pages} {1929} (\bibinfo {year} {2014})}\BibitemShut
  {NoStop}%
\bibitem [{\citenamefont {Battaglia}\ \emph {et~al.}(2018)\citenamefont
  {Battaglia}, \citenamefont {Hamrick}, \citenamefont {B.}, \citenamefont {S.},
  \citenamefont {Zambaldi}, \citenamefont {Malinowski}, \citenamefont
  {Tacchetti}, \citenamefont {Raposo}, \citenamefont {Santoro}, \citenamefont
  {Faulkner}, \citenamefont {Gulcehre}, \citenamefont {Song}, \citenamefont
  {Ballard}, \citenamefont {Gilmer}, \citenamefont {Dahl}, \citenamefont
  {Vaswani}, \citenamefont {Allen}, \citenamefont {Nash}, \citenamefont
  {Langston}, \citenamefont {Dyer}, \citenamefont {Heess}, \citenamefont
  {Wierstra}, \citenamefont {Kohli}, \citenamefont {Botvinick}, \citenamefont
  {Vinyals}, \citenamefont {Li},\ and\ \citenamefont {Pascanu}}]{GNN1}%
  \BibitemOpen
  \bibfield  {author} {\bibinfo {author} {\bibfnamefont {P.~W.}\ \bibnamefont
  {Battaglia}}, \bibinfo {author} {\bibfnamefont {J.~B.~C.}\ \bibnamefont
  {Hamrick}}, \bibinfo {author} {\bibfnamefont {V.}~\bibnamefont {B.}},
  \bibinfo {author} {\bibfnamefont {A.}~\bibnamefont {S.}}, \bibinfo {author}
  {\bibfnamefont {V.}~\bibnamefont {Zambaldi}}, \bibinfo {author}
  {\bibfnamefont {M.}~\bibnamefont {Malinowski}}, \bibinfo {author}
  {\bibfnamefont {A.}~\bibnamefont {Tacchetti}}, \bibinfo {author}
  {\bibfnamefont {D.}~\bibnamefont {Raposo}}, \bibinfo {author} {\bibfnamefont
  {A.}~\bibnamefont {Santoro}}, \bibinfo {author} {\bibfnamefont
  {R.}~\bibnamefont {Faulkner}}, \bibinfo {author} {\bibfnamefont
  {C.}~\bibnamefont {Gulcehre}}, \bibinfo {author} {\bibfnamefont
  {F.}~\bibnamefont {Song}}, \bibinfo {author} {\bibfnamefont {A.}~\bibnamefont
  {Ballard}}, \bibinfo {author} {\bibfnamefont {J.}~\bibnamefont {Gilmer}},
  \bibinfo {author} {\bibfnamefont {G.~E.}\ \bibnamefont {Dahl}}, \bibinfo
  {author} {\bibfnamefont {A.}~\bibnamefont {Vaswani}}, \bibinfo {author}
  {\bibfnamefont {K.}~\bibnamefont {Allen}}, \bibinfo {author} {\bibfnamefont
  {C.}~\bibnamefont {Nash}}, \bibinfo {author} {\bibfnamefont {V.~J.}\
  \bibnamefont {Langston}}, \bibinfo {author} {\bibfnamefont {C.}~\bibnamefont
  {Dyer}}, \bibinfo {author} {\bibfnamefont {N.}~\bibnamefont {Heess}},
  \bibinfo {author} {\bibfnamefont {D.}~\bibnamefont {Wierstra}}, \bibinfo
  {author} {\bibfnamefont {P.}~\bibnamefont {Kohli}}, \bibinfo {author}
  {\bibfnamefont {M.}~\bibnamefont {Botvinick}}, \bibinfo {author}
  {\bibfnamefont {O.}~\bibnamefont {Vinyals}}, \bibinfo {author} {\bibfnamefont
  {Y.}~\bibnamefont {Li}}, \ and\ \bibinfo {author} {\bibfnamefont
  {R.}~\bibnamefont {Pascanu}},\ }\href@noop {} {\enquote {\bibinfo {title}
  {Relational inductive biases, deep learning, and graph networks},}\ }
  (\bibinfo {year} {2018})\BibitemShut {NoStop}%
\bibitem [{\citenamefont {Battaglia}\ \emph {et~al.}(2016)\citenamefont
  {Battaglia}, \citenamefont {Pascanu}, \citenamefont {Lai}, \citenamefont
  {Jimenez~Rezende} \emph {et~al.}}]{GNN2}%
  \BibitemOpen
  \bibfield  {author} {\bibinfo {author} {\bibfnamefont {P.}~\bibnamefont
  {Battaglia}}, \bibinfo {author} {\bibfnamefont {R.}~\bibnamefont {Pascanu}},
  \bibinfo {author} {\bibfnamefont {M.}~\bibnamefont {Lai}}, \bibinfo {author}
  {\bibfnamefont {D.}~\bibnamefont {Jimenez~Rezende}},  \emph {et~al.},\
  }\bibfield  {title} {\enquote {\bibinfo {title} {Interaction networks for
  learning about objects, relations and physics},}\ }\href@noop {} {\bibfield
  {journal} {\bibinfo  {journal} {Advances in Neural Information Processing
  Systems}\ }\textbf {\bibinfo {volume} {29}},\ \bibinfo {pages} {4502}
  (\bibinfo {year} {2016})}\BibitemShut {NoStop}%
\bibitem [{\citenamefont {Scarselli}\ \emph {et~al.}(2009)\citenamefont
  {Scarselli}, \citenamefont {Gori}, \citenamefont {Tsoi}, \citenamefont
  {Hagenbuchner},\ and\ \citenamefont {Monfardini}}]{GNN3}%
  \BibitemOpen
  \bibfield  {author} {\bibinfo {author} {\bibfnamefont {F.}~\bibnamefont
  {Scarselli}}, \bibinfo {author} {\bibfnamefont {M.}~\bibnamefont {Gori}},
  \bibinfo {author} {\bibfnamefont {A.~C.}\ \bibnamefont {Tsoi}}, \bibinfo
  {author} {\bibfnamefont {M.}~\bibnamefont {Hagenbuchner}}, \ and\ \bibinfo
  {author} {\bibfnamefont {G.}~\bibnamefont {Monfardini}},\ }\bibfield  {title}
  {\enquote {\bibinfo {title} {The graph neural network model},}\ }\href@noop
  {} {\bibfield  {journal} {\bibinfo  {journal} {IEEE Transactions on Neural
  Networks}\ }\textbf {\bibinfo {volume} {20}},\ \bibinfo {pages} {61}
  (\bibinfo {year} {2009})}\BibitemShut {NoStop}%
\end{thebibliography}%

\section*{Acknowledgements}
The authors would like to thank Marjolein de Jager for many discussions.  L.F. and E.B. acknowledge funding from The Netherlands  Organisation  for  Scientific  Research  (NWO)  (Grant  No. 16DDS004), and L.F. acknowledges funding from NWO for a Vidi grant (Grant No. VI.VIDI.192.102). 

\end{document}